\documentclass[useAMS,usenatbib,usegraphicx]{mn2e}
\usepackage{times}

\newcommand{\lsim}{\lower0.6ex\vbox{\hbox{$ \buildrel{\textstyle <}\over{\sim}\ $}}}
\newcommand{\gsim}{\lower0.6ex\vbox{\hbox{$ \buildrel{\textstyle >}\over{\sim}\ $}}}
\newcommand{\beq}{\begin{equation}}
\newcommand{\eeq}{\end{equation}}

\title[Mass assembly history and infall time of the Fornax dwarf spheroidal galaxy]
{
Mass assembly history and infall time of the Fornax dwarf spheroidal galaxy
}

\author[M.-Y. Wang et al.]
{Mei-Yu Wang$^{1}$\thanks{Email:meiyu@physics.tamu.edu}, Louis E. Strigari$^{1}$, Mark R. Lovell$^{2,3}$, Carlos S. Frenk$^4$, and Andrew R. Zentner$^5$ \\
$^1$Department of Physics and Astronomy, Mitchell Institute for Fundamental Physics and Astronomy, Texas A\&M University, College Station, TX 77843-4242\\
$^2$GRAPPA, Universiteit van Amsterdam, Science Park 904, 1098 XH Amsterdam, The Netherlands\\
$^3$Instituut-Lorentz for Theoretical Physics, Niels Bohrweg 2, 2333 CA Leiden, The Netherlands\\
$^4$Institute for Computational Cosmology, Durham University, South Road, Durham, UK, DH1 3LE\\
$^5$Department of Physics $\&$ Astronomy, and Pittsburgh Particle physics, Astrophysics, and Cosmology Center (Pitt PACC), University of Pittsburgh,\\
 Pittsburgh, PA 15260 
}
\date{Released 2015 Xxxxx XX}

\pagerange{\pageref{firstpage}--\pageref{lastpage}} \pubyear{2015}

\begin{document}

\maketitle

\begin{abstract}
We use cosmological simulations to identify dark matter subhalo host candidates of the Fornax dwarf spheroidal galaxy using the stellar kinematic properties of Fornax. We consider cold dark matter (CDM), warm dark matter (WDM), and decaying dark matter (DDM) simulations for our models of structure formation. The subhalo candidates in CDM typically have smaller mass and higher concentrations at z = 0 than the corresponding candidates in WDM and DDM. We examine the formation histories of the $\sim$ 100 Fornax candidate subhalos identified in CDM simulations and, using approximate luminosity-mass relationships for subhalos, we find two of these subhalos that are consistent with {\em both} the Fornax luminosity and kinematics. These two subhalos have a peak mass over ten times larger than their z = 0 mass. We suggest that in CDM the dark matter halo hosting Fornax must have been severely stripped of mass and that it had an infall time into the Milky Way of $\sim$ 9 Gyr ago. In WDM, we find that candidate subhalos consistent with the properties of Fornax have a similar infall time and a similar degree of mass loss, while in DDM we find a later infall time of $\sim$ 3 $-$ 4 Gyr ago and significantly less mass loss. We discuss these results in the context of the Fornax star formation history, and show that these predicted subhalo infall times can be linked to different star formation quenching mechanisms. This emphasizes the links between the properties of the dark matter and the mechanisms that drive galaxy evolution.

\end{abstract} 

\section{Introduction}
\label{section:introduction}

The temperature anisotropies in the cosmic microwave background (CMB) anisotropy spectrum measured by the Wilkinson Microwave Anisotropy Probe (WMAP) \citep{WMAP9} and PLANCK \citep{Planck_13}, and observations of the large-scale ($k \lsim 0.1$ Mpc $h^{-1}$) galaxy clustering spectrum measured by the 2dF Galaxy Redshift Survey \citep{Cole_etal05} and Sloan Digital Sky Survey \citep{Tegmark_etal06} have shown that the large-scale structure formation is consistent with the $\Lambda$ cold dark matter (CDM) model~\citep{Frenk:2012ph}. The observation of the CDM component implies physics beyond the standard model, and many dark matter candidates exist within extensions to the standard model of particle physics that behave as CDM~\citep{jungman_etal96b}. Though CDM is theoretically well-motivated, there are both theoretical (e.g.~\cite{Abazajian:2012ys,Zurek:2013wia}) and observational~\citep{Weinberg:2013aya,Bulbul2014,Boyarsky2014,Boyarsky_etal14b} interests in considering alternatives. In fact, a broad exploration of particle dark matter candidates finds that many viable models behave differently than CDM, particularly on small scales, implying that observations of dark matter structure on small scales may provide a unique test of different particle dark matter candidates. 

Aside from the Magellanic Clouds, the eight largest satellite galaxies of the Milky Way are dwarf spheroidal (dSph) galaxies and the internal kinematics of these dSphs offer one of the best prospects for understanding the properties of particle dark matter from small scale astronomical observations (for a recent review see~\cite{Walker:2012td}). Stellar kinematics unambiguously indicate that the dSphs are dark matter-dominated~\citep{Walker:2007ju}, and their measured potentials have been used to determine whether their dark matter profiles are consistent with an NFW density profile long predicted by CDM~\citep{Navarro_etal96,navarro_etal97}. However, in spite of the high quality data sets, at present it is unclear whether the data indicate that the dark matter distributions in dSphs are in conflict with the NFW model~\citep{Walker_etal11, Amorisco_etal13} or are consistent with it~\citep{Jardel:2012am,Breddels:2013qqh,Strigari_etal14}. 

Improvements in N-body simulations and in hydrodynamic simulations of Milky Way-like dark matter halos and their corresponding populations of subhalos now provide even more detailed predictions for the dark matter distributions of the dSphs. Detailed fitting of the stellar kinematics and photometry to subhalos in CDM N-body simulations indicate that the dSphs reside in dark matter halos with maximum circular velocity of approximately 20-25 km/s~\citep{Strigari_etal10}. More general fits to the dSph stellar kinematics in alternative dark matter model simulations such as warm dark matter (WDM)~\citep{Lovell_etal14}, decaying dark matter (DDM) ~\citep{Wang_etal14}, and self-interacting dark matter (SIDM)~\citep{Vogelsberger_etal12} indicate that the maximum circular velocities are larger than in CDM~\citep{Boylan-Kolchin_etal11}. Recent CDM hydrodynamic simulations find that these dark matter-only simulations neglect the important effect of baryons, which modify the $z=0$ maximum circular velocities of dSphs by about 15$\%$~\citep{Zolotov_etal12,Brooks_etal13,Brooks_etal14,Sawala_etal14,Sawala:2014xka}. 

Identifying subhalos that are consistent with hosting dSphs in CDM as well in alternative dark matter scenarios can shed light on the cosmological evolution of those subhalos and, perhaps, the galaxies that they contain. Such identifications may pave the way for the development of specific predictions of both CDM and alternative models that can serve as true tests of the models. Exploiting stellar kinematics for this purpose is complementary to and significantly more robust than using predictions for the luminosities of galaxies within subhalos, because these predicted luminosities are extremely uncertain and rely on extrapolating phenomenological scaling relations outside of their established domain~\citep{Garrison-Kimmel:2014vqa}. It is additionally complementary to methods that utilize measurements of the orbital motions of Milky Way satellites~\citep{Rocha_etal12b,Sohn_etal13,Kallivayalil:2013xb}.

In this paper we identify dark matter subhalo host candidates of the Fornax dSph in CDM, WDM, and DDM N-body simulations by matching to the observed kinematics and photometry. The non-CDM based models that we study exhibit DM free-streaming effects that are not present in CDM. For WDM models, previous studies suggest that an equivalent thermal relic mass $\sim$ keV generates a truncation of the matter power spectrum on scales $\sim$ a few Mpc~\citep{Bode_etal01}. This effect suppresses the formation of small structure below the WDM free-streaming scale, resulting in delayed formation of halos.~\cite{Lovell_etal14} use galactic zoom-in simulations to show that typical galactic subhalos are less concentrated than their CDM counterparts because they form at later times. Sterile neutrinos are a canonical WDM candidate, and the decays of sterile neutrinos provide a possible origin for the detection of an unexplained X-ray line observed at 3.55 keV in the Galactic Centre, M31, and galaxy clusters ~\citep{Bulbul2014,Boyarsky2014,Boyarsky_etal14b}. In DDM models, the DM free-streaming is delayed since the excess velocity imparted from the DM decay is introduced with a lifetime comparable or greater to Hubble time. This can avoid the tight limits placed by high-redshift phenomena like Lyman-$\alpha$ forest \citep{Wang_etal13} and can impact galactic substructure~\citep{Wang_etal14}. In DDM, at high redshift the structure formation is similar to CDM until the age of the Universe is comparable to decay lifetime~\citep{Wang_etal12}.

We focus on the stellar kinematics of Fornax because it has a high-quality kinematic data sample and the dark matter potentials of subhalos in all of our simulations are well-resolved on the scale of the Fornax half-light radius. With the candidate host subhalos of Fornax identified, we determine the dynamical properties of the host subhalos, such as the present day mass, the peak mass, and the maximum circular velocity. With candidate host subhalos identified in each simulation, we examine the assembly histories of the Fornax host candidates. To assign luminosity to our Fornax candidates we use a simple relationship between the stellar mass and peak halo mass, and from this we identify Fornax subhalo candidates in CDM, WDM, and DDM that are consistent with both its kinematics and luminosity. These Fornax candidates allow us to predict the infall times and degree of tidal stripping of these subhalos, and we use these quantities to connect to models for the Fornax star formation history in each simulation. 

\begin{figure}
\includegraphics[height=11.8cm]{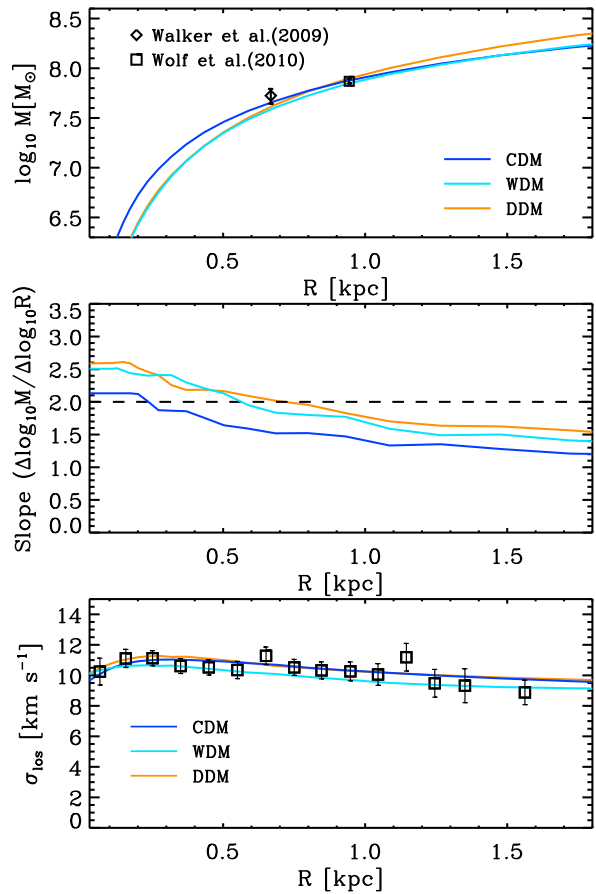}
\caption{ 
\textit{Top}: DM distribution of subhalos selected from three simulations with different DM models. The navy line represents a subhalo drawn from the Aquarius simulation (CDM), the aqua line is from the WDM simulation with WDM thermal relic mass = 2.3 keV, and the orange line is from the DDM simulations. These three subhalos provide good-fits to the Fornax stellar kinematic profile and photometry data. Mass estimations for Fornax from previous studies are also shown. The squared data points with 1 $\sigma$ error bars show the mass estimation at the 3D half-light radius from \citet{Wolf_etal10}, and the diamond points with 1 $\sigma$ error bars from \citet{Walker_etal09}. \textit{Middle}: Slopes of logarithmic mass profiles from these three subhalos. The horizontal dash line marks where slope =2, which is the theoretical predicted slope for NFW profiles as R $\to$ 0. The non-CDM subhalos have slope $>$ 2.0 in the inner region. \textit{Bottom}: The best-fit line-of-sight velocity dispersion derived using the jeans equation formalism with $\beta$ = constant model.  
}
\label{fig:mr}
\end{figure}
 
The outline of the paper is as follows. In \S~\ref{section:simulations}, we briefly describe the properties of simulations used in our analysis. In~\S~\ref{section:fitting} we review our procedures for identifying Fornax subhalo candidates using its stellar kinematic data and photometry data. In~\S~\ref{section:results} we present our results for the subhalo assembly histories, and discuss how these quantities can be used to determine the Fornax infall time, star formation history, and quenching mechanisms. Lastly we draw our conclusions in \S~\ref{section:conclusion}.

\section{Simulations}
\label{section:simulations}

In this section we briefly describe the cosmological simulations that we utilize. For more details we refer to the original simulation papers (CDM : \cite{Springel_etal08}, WDM : \cite{Lovell_etal14}, and DDM : \cite{Wang_etal14}).

\begin{figure*}
\includegraphics[height=4.6cm]{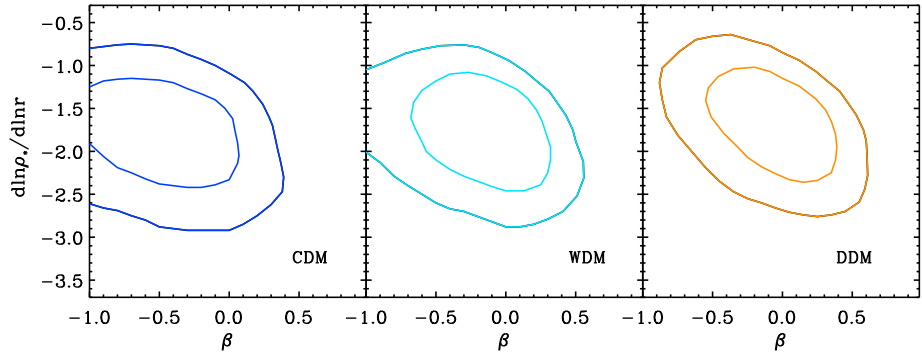}
\caption{ The 68$\%$ and 95$\%$ contour region for the stellar density slope at r=600 pc versus velocity anisotropy $\beta$. From left to right, the contours are drawn from the MCMC results of the Jeans equation fitting for the subhalos shown in Figure~\ref{fig:mr} (left: CDM, middle: WDM, right: DDM).
}
\label{fig:contour}
\end{figure*}

For the CDM model, we utilize the Aquarius simulations, which are six realizations (Aquarius A to F) of galactic zoom-in simulations~\citep{Springel_etal08}. These simulations are generated using the Gadget code~\citep{Springel_etal08} and use cosmological parameters consistent with the one-year and five-year~\textit{Wilkinson Microwave Anisotropy Probe} (WMAP) data: $H_0$ = 73 km/s/Mpc, $\Omega_m$ = 0.25, $\Omega_{\Lambda}$ = 0.75, $\sigma_8$ = 0.9, and $n_s$ = 1. We adopt the level-2 resolution simulations as our main sample (with Plummer gravitational softening length $\epsilon$ = 65.8 pc and particle mass $m_p$ = 1.399$\times10^4M_{\odot}$$-$6.447$\times10^3M_{\odot}$), and utilize the highest level-1 resolution for Aquarius A (with $\epsilon$ = 20.5 pc and particle mass $m_p$ = 1.712$\times10^3M_{\odot}$) to perform resolution tests to understand the effects of the force softening scale on our analysis. The details of these tests are described in the Appendix~\S~\ref{Resolution}. The properties of the Aquarius simulations are shown in Table~\ref{tb:simulation}.

For the WDM models, we use the simulations described in~\cite{Lovell_etal14}. To account for WDM physics,~\cite{Lovell_etal14} re-simulate the Aquarius A halo using initial condition wave amplitudes that are rescaled with thermal relic WDM power spectra from~\cite{Viel_etal05}. We adopt their ``high resolution" suite that corresponds to level-2 in the original Aquarius notation with $m_p$ = 1.55 $\times10^4 M_{\odot}$ and $\epsilon$ = 68.1 pc. This suite includes two WDM simulations with equivalent thermal relic masses of $2.3$ keV and $1.6$ keV, and their CDM counterpart simulations. The $2.3$ keV thermal relic is a good approximation to the matter power spectrum of a 7 keV sterile neutrino that is resonantly produced in a lepton asymmetry $L_6\sim$20, which translates to a transfer function warmer than that inferred from the 3.55 keV line ~\citep{Venumadhav15, Lovell_etal15}. We take $1.6$ keV model as a case that is ruled out by current Lyman-$\alpha$ forest limits (e.g. ~\citep{Viel_etal13}) and $2.3$ keV as the model likely allowed by such limits for comparison. The cosmological parameters are derived from WMAP7: $H_0$ = 70.4 km/s/Mpc, $\Omega_m$ = 0.272, $\Omega_{\Lambda}$ = 0.728, $\sigma_8$ = 0.81, and $n_s$ = 0.967.  Note that this is different than the original Aquarius simulations for which WMAP1 cosmology was implemented. We describe the effects of the different cosmology on our results in Appendix~\S~\ref{w1w7}. The self-bound halos in Aquarius simulations and WDM simulations were identified using the \textsc{SUBFIND} algorithm~\citep{Springel_etal01}.

For our DDM model, we use the set of simulations that implement late-time DDM physics from~\cite{Wang_etal14}. In these DDM models, a dark matter particle of mass $M$ decays into a less massive daughter particle of mass $m = (1-f)M$ with $f \ll 1$ and a significantly lighter, relativistic particle, with a lifetime $\Gamma^{-1}$, where $\Gamma$ is the decay rate. The stable daughter particle acquires a recoil kick velocity, $V_{k}$, the magnitude of which depends upon the mass splitting between the decaying particle and the daughter particle. The DDM simulations are generated using a modified version of ~\textsc{GADGET-3}~\citep{peter_etal10}. Here we consider the case with decay lifetime $\Gamma^{-1}$= 10 Gyr and kick velocity $V_{k}$ = 20 km/s. This model has been show to have interesting implications for dark matter small-scale structure~\citep{Wang_etal14} and is allowed by current Lyman-$\alpha$ forest limits~\citep{Wang_etal13}. The cosmology used is based on WMAP7 results with $H_0$=71 km/s/Mpc, $\Omega_m$=0.266, $\Omega_{\Lambda}$=0.734, $\sigma_8$=0.801, and $n_s$=0.963. We use \textsc{Amiga Halo Finder (AHF)} \citep{Knollmann_etal09} for halo finding and the merger tree is constructed using~\textsc{Consistent Trees}~\citep{Behroozi_etal13b}. 

We define the virial mass  $(M_{200b})$ of the galactic halo as the mass enclosed within the region of 200 times the background for all simulations, which corresponds to the $M_{50}$ in~\citet{Springel_etal08}. We adopt this definition for the same reason quoted in~\citet{Springel_etal08}, namely that it yields the largest radius among other conventional halo definition and hence the largest number of substructures.

\begin{table*}
{\renewcommand{\arraystretch}{1.3}
\renewcommand{\tabcolsep}{0.2cm}
\begin{tabular}{l c c c c c c c}
\hline 
\hline
Simulations &  Particle mass $m_p$& Force Softening $\epsilon$ & $M_{200b}$ & $r_{200b}$ &Dark Matter Properties\\
  & [$M_{\odot}$] &  [pc] &  [$M_{\odot}$] &  [kpc] & \\
\hline
Aq-A1 &1.712$\times 10^3$&20.5 &2.52$\times 10^{12}$ & 433.5 &CDM\\
Aq-A2 &1.370$\times 10^4$&65.8 &2.52$\times 10^{12}$& 433.5 &CDM\\
Aq-B2 &6.447$\times 10^3$&65.8 &1.05$\times 10^{12}$& 323.1 &CDM\\
Aq-C2 &1.399$\times 10^4$&65.8 &2.25$\times 10^{12}$& 417.1 &CDM\\
Aq-D2 &1.397$\times 10^4$&65.8 &2.52$\times 10^{12}$& 433.2 &CDM\\
Aq-E2 &9.593$\times 10^3$&65.8 &1.55$\times 10^{12}$& 368.3 &CDM\\
Aq-F2 &6.776$\times 10^3$&65.8 &1.52$\times 10^{12}$& 365.9 &CDM\\
\hline
Aq-A2 w7 &1.545$\times 10^4$&68.2 &2.53$\times 10^{12}$ & 432.1&CDM\\
Aq-A2-$m_{1.6}$ &1.545$\times 10^4$&68.2 &2.49$\times 10^{12}$& 429.9 &WDM ($m_{\mathrm{\small{WDM}}}$=1.6keV)\\
Aq-A2-$m_{2.3}$ &1.545$\times 10^4$&68.2 &2.52$\times 10^{12}$& 431.4 &WDM ($m_{\mathrm{\small{WDM}}}$=2.3keV)\\
\hline
Z13-CDM &2.40$\times 10^4$ &72.0 &1.31$\times 10^{12}$ & 335.2 &CDM\\
Z13-t10-v20 & 2.40$\times 10^4$ &72.0 &1.16$\times 10^{12} $ & 336.2 &DDM $(\Gamma^{-1}$=10 Gyr, $V_k$=20.0 km/s)\\
\end{tabular}
\medskip
\caption{Parameters of simulations. The mass of the galactic halo $M_{200b}$ is defined as the mass enclosed within the region of 200 times of the background density. }
 }
 \label{tb:simulation}
\end{table*}

\section{Fitting subhalos to stellar kinematics}
\label{section:fitting}
In this section we discuss our method for fitting subhalos in simulations to the stellar kinematic and photometric data of Fornax. For our theoretical analysis we use the spherical jeans equation, and allow for a constant but non-zero anisotropic stellar velocity dispersion. We utilize this theory in order to efficiently identify a large sample of Fornax candidates. For comparison studies with similar motivations have been undertaken~\citep{Boylan-Kolchin_etal11,Garrison-Kimmel:2014vqa} that have used the mass estimator of~\citet{Walker_etal09} and \citet{Wolf_etal10} to estimate maximum circular velocities for the dSphs. As we show below our results are in good agreement with these previous results. Since we start at the level of the jeans equation, we are able to identify preferred orbital structure of the stars given the underlying subhalo potentials. 

We assume that the potential is spherically symmetric, dispersion-supported, and in dynamical equilibrium, so that we can derive the stellar line-of-sight velocity dispersion profile as a function of projected radius $R$~\citep{Binney_etal82}:
\beq
\sigma^2_{los}(R)={2 \over I_{\ast}(R)} \int^{\infty}_{R}[1-\beta(r){R^2 \over r^2}] {\rho_{\ast}(r)\sigma_r^2r \over \sqrt{r^2-R^2}} dr
\label{eq:vd_los}
\eeq
with
\beq
 I_{\ast}(R)=2 \int^{\infty}_{R} {\rho_{\ast}(r)r \over \sqrt{r^2-R^2}} dr.
 \label{eq:Abel}
\eeq
Here $\rho_{\ast}(r)$ is the 3D stellar density profile and $ I_{\ast}(R)$ is its 2D projection. 
The velocity anisotropy parameter is $\beta(r)\equiv$ 1-$\sigma^2_t(r)/\sigma^2_r(r)$, where $\sigma_r(r)$ is the radial velocity dispersion of stars and $\sigma_t(r)$ is the tangential velocity dispersion. These quantities satisfy the Jeans equation \citep{Binney_etal08}:
\beq
r{d(\rho_*\sigma^2_r) \over dr}+2\beta(r)\rho_*\sigma^2_r=-\rho_*(r){GM(<r)\over r}.
\label{eq:Jeans}
\eeq
For the case of constant non-zero $\beta$, the solution of Eq.~\ref{eq:Jeans} has a simple form:
\beq
\rho_{\ast}\sigma_r^2(r)={1 \over r^{2\beta}}\int^{\infty}_{r} {\rho_{\ast}(r')GM(<r') r'^{2\beta} \over r'^2} dr'. 
\label{eq:rho_sigma}
\eeq 

To model the 3D stellar density profile $\rho_{*}(r)$ as a function of radius we use this general form:
\beq
\rho_{\ast}(r) \propto {1 \over x^{a}(1+x^b)^{(c-a)/b}}, 
\label{eq:sp}
\eeq 
where x=$r$/$r_0$ and $a, b, c$ are free parameters that capture the stellar distribution slopes over different radii. A density profile of this form has been found to adequately describe the photometry of the classical dSphs~\citep{Strigari_etal10}. Though the conversion of 3D to 2D profile is not a one-to-one relation, implying that different choice of parameters for the 3D profiles can provide similar 2D projected profiles, the present photometry data do give good constraints on the Fornax stellar distributions with profiles of the form Eq.~\ref{eq:sp}. We fix the normalization of stellar mass by assuming mass-to-light ration $M_*$/L $= 1$ and adopt the Fornax V-band luminosity value of $L_v$=1.7$\times 10^{7} M_{\odot}$~\citep{McConnachie2012}. 

With the above equations we utilize the following algorithm to model the Fornax line-of-sight velocity dispersion given the subhalos in each of our simulations. We begin by finding subhalos in our galactic zoom-in simulations using halo finders, and for each subhalo determine the dark matter distribution and thus the potential as a function of radius. Given the potential of each subhalo, the stellar distribution described by the parameters $a,b,c,r_0$, and velocity anisotropy by $\beta$, we solve the jeans equation to determine the line-of-sight velocity dispersion. We then fit to the photometric and kinematic data by marginalizing over these parameters $a,b,c,r_0$, and $\beta$ via a Markov Chain Monte Carlo (MCMC) method to determine the best fitting value for the model parameters. 

For our fits to the Fornax stellar kinematics we define
\beq
\chi^2_{\sigma} = \sum^{N_{bins}}_{i=1} {[\hat{\sigma}_i-\sigma_{los}(R_i)]^2 \over \epsilon^2_i},
\label{eq:chi_vd}
\eeq
where $N_{bins}$ is the number of annuli, $R_i$ is the mean value of the projected radius of stars in the $i^{th}$ annulus, and $\sigma_{los}(R_{i}) $ is the derived velocity dispersion for a given subhalo in the $i^{th}$ annulus. The line-of-sight velocity dispersion from the binned data is $\hat{\sigma_i}$ and the corresponding error is $\epsilon_i$ in each annulus.   

The kinematic datasets that we use consist of line-of-sight stellar velocities from the samples of~\cite{Walker_etal09b}. We consider only stars with $>$ 90$\%$ probability of membership, which gives us a sample of 2409 Fornax member stars. We bin the velocity data in circular annuli and derive the mean line-of-sight velocity in each annulus as function of $R_i$. We calculate the line-of-sight velocity dispersion $\hat{\sigma}$ and their error $\epsilon$ in each annulus following the method described in~\cite{Strigari_etal10}.

For our fits to the Fornax photometry we define  
\beq
\chi^2_{IR} = \sum^{N_{IR}}_{i=1} {[\hat{I_*}(R_i)-I_*(R_i)]^2 \over \epsilon^2_{IR, i}},
\label{eq:chi_IR}
\eeq
where $N_{IR}$ is the number of radial bins, $R_i$ is the radius of the $i^{th}$ data point, and $\hat{I_*}(R_i) $ is the derived 2D stellar surface density in the $i^{th}$ radius bin and $ \epsilon^2_{IR, i}$ is the uncertainty. For our photometric data we use the measurements from~\cite{Coleman_etal05}. Note that in our analysis we assume that Fornax contains a single population of stars.

For our MCMC, we adopt uniform priors over the following range: 0 $\leq$ a $\leq$ 2, 0.5 $\leq$ b $\leq$ 5, 4 $\leq$ c $\leq$ 8, 0.5 $\leq$ $r_0$ $\leq$ 2, -1.0 $\leq$ $\beta$ $\leq$ 1.0. The MCMC calculation is performed using a modified version of the publicly available code \textsc{CosmoMC} \citep{Lewis_etal02} as an MCMC engine with our own likelihood functions. We adopt a simple form of the likelihood function as the summation of the $\chi^2$ from both the stellar kinematic data and photometry over their degree of freedom: $\chi^2_{tot}/d.o.f.$ = $\chi^2_{\sigma}/d.o.f.$ + $\chi^2_{IR}/d.o.f.$, where $\chi_{\sigma}^2$ and $\chi_{IR}^2$ are calculated using Eq.~\ref{eq:chi_vd} and Eq.~\ref{eq:chi_IR} with degree of freedom (\textit{d.o.f.}) equal to number of data bins minus the number of free parameters plus one. We have 15 radial bins in the velocity dispersion data and 19 radial bins in the photometry light profile data. We then select the subhalos that satisfy the condition of $\chi^2_{tot}/d.o.f. \leq 3.0$ as good-fits to the Fornax kinematic and photometry data. We choose these criteria in order to obtain a conservatively large subhalo sample of candidates that can host Fornax.

\begin{figure*}
\includegraphics[height=5.8cm]{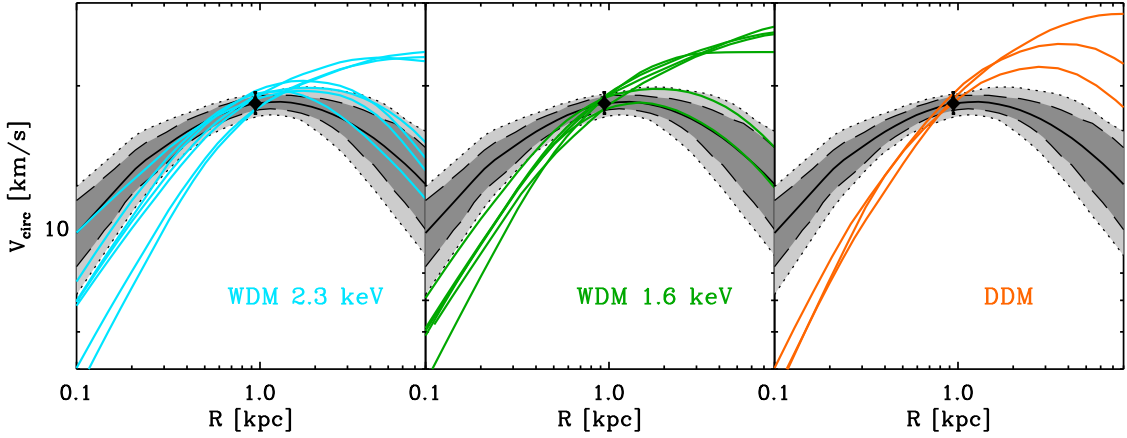}
\caption{ Circular velocity curves of subhalos that are good fits to the Fornax stellar kinematic and photometry data. Notice that for each curve the stellar mass contribution has been included. The dark gray shaded areas include 68$\%$ of the subhalo curves from Aquarius simulation A to F, and the light gray areas are for 95 $\%$ of the subhalos. The solid black lines are the average of all the Aquarius CDM fits. From left to right panel, the colored lines show the subhalos from the WDM simulation with 2.3 keV thermal relic mass (left panel, aqua), the WDM simulation with 1.6 keV thermal relic mass (middle panel, green), and the DDM simulation with lifetime =10 Gyr and kick velocity = 20 km/s (right panel, orange)
}
\label{fig:vcir}
\end{figure*}

\section{Results}
\label{section:results}
We now present the results of our analysis. We begin by presenting the properties of the dark matter subhalos in the different simulations that are candidate Fornax hosts using stellar kinematics alone. We then combine with a simple model for subhalo luminosities to identify subhalos that match both the luminosity and kinematics of Fornax, and present results for the assembly history of these subhalos. 

\subsection{Properties of Fornax Candidate Host Subhalos at z=0}
In this subsection we discuss the properties of Fornax candidate host subhalos at $z=0$. We begin by discussing the density profiles. for which we compare the central densities of the subhalos from the different simulations. We then discuss the total subhalo masses and maximum circular velocities of the host subhalo candidates. 

\label{subsection: Best-fit subhalos}

\subsubsection{Subhalo density profiles and correlations}
We begin by examining the density profiles of the subhalos that provide good fits to the photometry and line-of-sight velocity dispersion profiles. Figure~\ref{fig:mr} shows the line-of-sight velocity dispersions for three different representative subhalos, one from each  of the CDM, WDM, and DDM simulation. For CDM, here we choose the subhalo that has the minimum value of $\chi^2_{tot}/d.o.f.$, while for DDM and WDM we choose the subhalos that have $\chi^2_{tot}/d.o.f.$ $<$ 3 and have the shallowest central density profiles. For WDM (DDM), the $\chi^2_{tot}/d.o.f.$ is specifically 1.67 (0.85), and the p-value for the velocity dispersion fit is 0.4 (0.84). For the CDM subhalo, the $\chi^2_{tot}/d.o.f.$ is 0.93 while the p-value for the velocity dispersion fit is 0.8. Thus in all three cases the fits are good from a statistical standpoint .

Figure~\ref{fig:mr} also shows both the mass distribution and the log-slopes as a function of radius of the selected subhalos. The CDM subhalo is the most centrally-concentrated at radii $\lsim$ 500 pc, generating a central slope of $\Delta log_{10}M/\Delta log_{10} R$ $\sim$ 2, which is consistent with the predicted value from an NFW profile. The DDM subhalo has a central slope of $\sim$ 2.6, while the WDM subhalo has a central slope of $\sim$ 2.5. Thus the WDM subhalos have ``soft" cores, which are expected since WDM subhalos are less concentrated than their CDM counterparts because they form at a later time~ \citep{Lovell_etal12}. The DDM subhalo has a slightly shallower core-like feature than in WDM because the effects of dark matter decay are scale-dependent~\citep{Wang_etal14}. 

As Figure~\ref{fig:mr} indicates, the predicted line-of-sight velocity dispersions are very similar, even though the best fitting values of $a,b,c,r_0$ and $\beta$ are different in each case. This highlights an explicit degeneracy between the dark matter potential, the velocity anisotropy, and the stellar density profile. At the level of a jeans-based analysis, the degeneracy between the dark matter potential and the velocity anisotropy has been well-studied and has been known to preclude determination of the dark matter profile shape~\citep{Strigari:2007vn}. 

The degeneracy between the stellar density profile and both the dark matter potential and the velocity anisotropy is less well-understood (see however~\cite{Strigari_etal10,Strigari_etal14}), so it is interesting to examine this further within the context of a Jeans-based analysis. Figure~\ref{fig:contour} highlights this degeneracy for the case of a single subhalo by showing the correlation between the stellar density slope at $r = 600$ pc versus anisotropy parameter. For all models, within 68$\%$ contour region $\beta$ spans a wide range and is degenerate with the stellar density concentration. The specific allowed regions change in each of the models, because they predict different DM density slopes.

\begin{figure*}
\includegraphics[height=8.3cm]{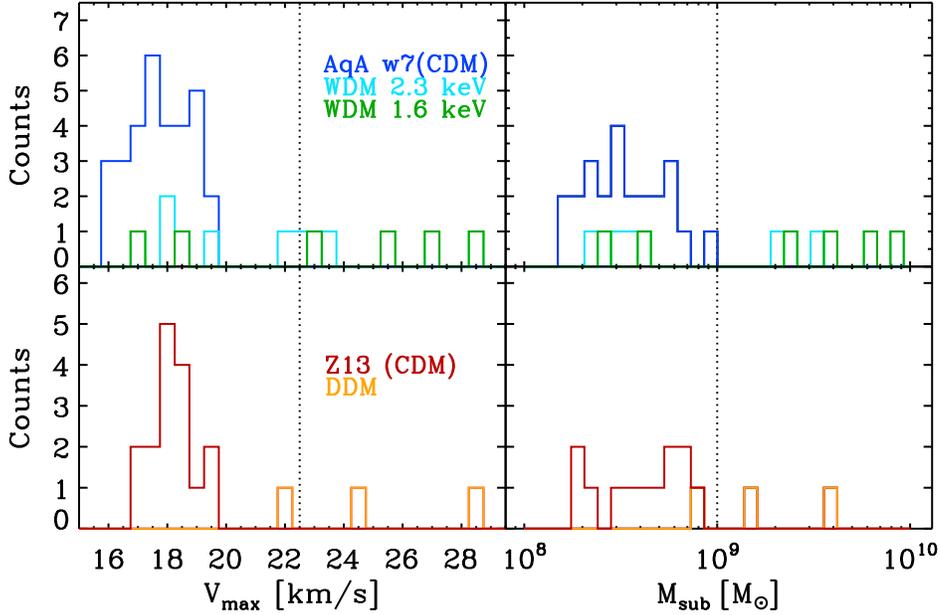}
\centering
\caption{Histograms of subhalo properties for the Fornax good-fits. Here we show two subhalo present-time properties: maximum circular velocity ($V_{max}$, left panels), and subhalo mass ($M_{sub}$, right panels). In the upper two panels the navy histograms show the Aquarius A results, the aqua ones are for the WDM 2.3 keV simulation, and the green ones are for the WDM 1.6 keV simulation. In the lower two panels the dark red histograms show the Z13 CDM results, and the orange ones are for the DDM simulation. We note that for the same galactic halo realization, the alternative DM scenarios presented here (WDM, DDM) generate much less substrucure than their CDM couterparts (see text in \S~\ref{subsubsection: Subhalo masses}).
}
\label{fig:param}
\end{figure*}

\begin{table}
\caption{Average subhalo properties of Fornax kinematic good-fits }
{\renewcommand{\arraystretch}{1.3}
\renewcommand{\tabcolsep}{0.2cm}
\begin{tabular}{l c c c c c c c}
\hline 
\hline
Simulations & $M_{sub}$ & $M_{peak}$ & $V_{max}$ \\
  & [$M_{\odot}$] & [$M_{\odot}$] & [km/s] \\
\hline
Aq-A2-w7 & 3.72$\pm$1.91$\times 10^8$& 1.55$\pm$2.01$\times 10^9$&17.70$\pm$1.02\\
Aq-A2-$m_{2.3}$&1.30$\pm$1.32$\times 10^9$&4.23$\pm$2.09$\times 10^9$&20.31$\pm$2.26\\
Aq-A2-$m_{1.6}$&3.52$\pm$3.15$\times 10^9$&5.05$\pm$2.67$\times 10^9$&23.22$\pm$4.57\\
\hline
\hline
Z13-CDM &4.50$\pm$2.00$\times 10^8$& 8.77$\pm$2.94$\times 10^8$&18.26$\pm$0.71\\
Z13-t10-v20&2.11$\pm$1.78$\times 10^9$& 4.87$\pm$5.10$\times 10^9$ &25.03$\pm$3.18\\
\hline
Aq-A2 &3.46$\pm$1.27$\times 10^8$&1.60$\pm$2.14$\times 10^9$&17.69$\pm$0.80\\
Aq-B2 &2.34$\pm$1.06$\times 10^8$&1.04$\pm$1.76$\times 10^9$&17.78$\pm$0.98\\
Aq-C2 &2.60$\pm$1.10$\times 10^8$&0.91$\pm$1.25$\times 10^9$&17.61$\pm$0.44\\
Aq-D2 &3.40$\pm$2.30$\times 10^8$&8.28$\pm$4.23$\times 10^8$&17.81$\pm$1.13\\
Aq-E2 &2.84$\pm$1.23$\times 10^8$&6.62$\pm$3.79$\times 10^8$&17.60$\pm$0.97\\
Aq-F2 &3.58$\pm$2.43$\times 10^8$&1.00$\pm$0.59$\times 10^9$&17.98$\pm$1.04\\
\hline
Aq Average &3.07$\pm$1.77$\times 10^8$&1.02$\pm$1.29$\times 10^9$&17.76$\pm$0.94\\
Aq Max &1.18$\times 10^9$&8.83$\times 10^9$&21.71\\
Aq Min &1.09$\times 10^8$&3.08$\times 10^8$&15.79\\
\hline
\end{tabular}
\medskip
\\
 }
\label{tb:summary}
\end{table}

\subsubsection{Subhalo masses and circular velocities} 
\label{subsubsection: Subhalo masses}

Though the density profiles of the CDM, WDM, and DDM subhalos are indistinguishable from the photometric and the kinematic data, the subhalos are distinct when considering more global properties. In Figure~\ref{fig:vcir} we show the circular velocity curves, $V_{circ}=\sqrt{G M(<r)/r}$, of subhalos that provide good fits to the Fornax kinematic data and photometry data. Note that for each curve the stellar mass contribution has been included, so that at a radius of 0.9 kpc, the enclosed mass contribution from stars is about 10$\%$ of the DM mass. The circular velocity curves in the WDM and DDM models are shallower in the center than the CDM subhalos, which is consistent with the discussion above on the individual best-fitting density profiles. Further, the WDM and DDM subhalos have larger maximum value, $V_{max}$, which is consistent with previous analyses~\citep{Lovell_etal12,Wang_etal14}. 

\begin{figure*}
\includegraphics[height=5.8cm]{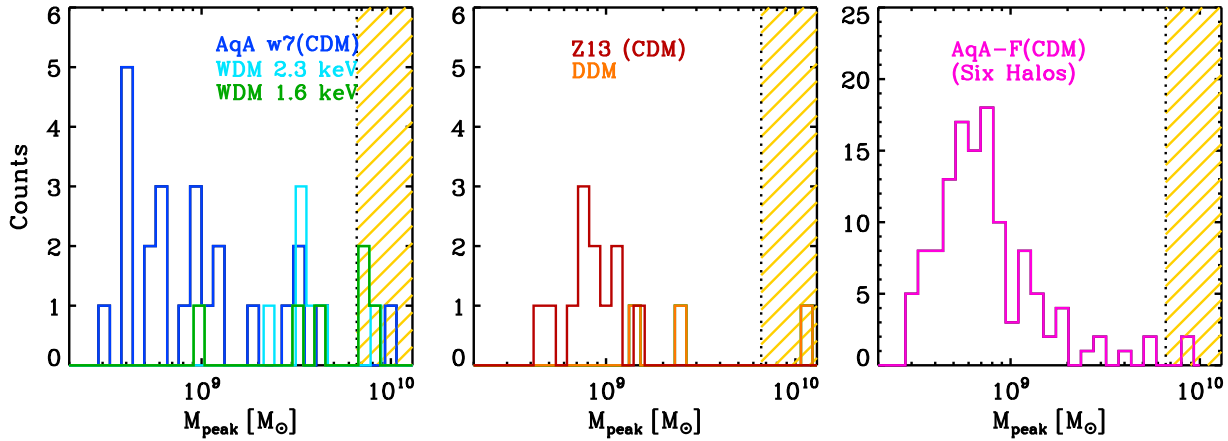}
\caption{ 
The  $M_{peak}$ distribution of the Fornax good-fit subhalos. The color notation in the left panel is the same as previous figures, and the right panel shows the distribution from all six Aquarius level-2 simulations (A-F). We note that there is only one galactic halo realization in the left and middle panel, while there are six in the left panel. The yellow shaded areas indicate the region where the luminosity of these objects are consistent with the Fornax luminosity using abundance matching methods from \citet{Behroozi_etal13}.
}
\label{fig:mpeak}
\end{figure*}

Figure~\ref{fig:param} shows the distribution of $V_{max}$ and the total mass of the subhalo, $M_{sub}$, for the CDM subhalo host candidates that are good fits to Fornax. Here we show only results from the CDM simulations Aq-A2-w7 and Z13, because from these simulations we are able to identify the subhalo counterparts from the WDM and DDM simulations, respectively. From Figure~\ref{fig:param} and also Table~\ref{tb:summary}, we show that CDM host subhalo candidates have $10^{8} M_{\odot} \le M_{sub}\le 10^{9} M_{\odot}$, $16 \, {\rm km/s} \, \le V_{max}\le 22 $ km/s. Table~\ref{tb:summary} shows the average $M_{sub}$ and $V_{max}$ of Fornax candidates from all our simulations. As is indicated, Aq-A2-w7 and Z13 provide a good representation of our entire sample of CDM simulations. The $V_{max}$ range of candidates from six Aquarius simulations (see Table~\ref{tb:summary}) have an average $\sim$17 km/s and an associated small variance, which agree well with results from previous study~\citep{Boylan-Kolchin_etal12}.

For comparison Figure~\ref{fig:param} shows the WDM and DDM subhalos that match the Fornax kinematics and photometry. In the WDM simulation with 2.3 keV WDM mass we identify seven host subhalo candidates with $M_{sub} \sim 0.3-3\times10^9 M_{\odot}$ and $V_{max} \sim$ 18-23 km/s. In the WDM simulation with 1.6 keV particle mass, we identify six host subhalo candidates with $M_{sub} \sim 0.2-8\times10^9 M_{\odot}$ and $V_{max} \sim$ 17-28 km/s. In the DDM simulation we identify three host subhalo candidates with $M_{sub} \sim 0.8-4\times10^9 M_{\odot}$ and $V_{max} \sim$ 22-28 km/s. These results indicate that from our Jeans-based modeling WDM and DDM predict a population of massive Fornax host subhalos that have shallower central dark matter density profiles. We note that alternative DM scenarios always predict much fewer Fornax candidates than their CDM counterpart simulations, simply because the number of subhalos are reduced significantly due to different DM properties. For example for the WDM case, Fig. 11 in~\cite{Lovell_etal14} shows that the numbers of subhalos with $V_{max} \ge$ 15 km/s are 14 (1.6 keV), 28 (2.3 keV) and 120 (CDM). For the DDM case the subhalo number with $V_{max} \ge$ 15 km/s is 16 and 56 for the CDM counterpart. The subhalo numbers in different CDM halos reflect mainly the differences in halo mass and merger history.

\subsection{Predictions for Luminosity and Effects of Reionization}
\label{subsection: luminosity}
To this point we have not used the luminosity of Fornax as a constraint, other than to consider its impact on the stellar kinematics and photometry. In doing so we have implicitly assumed that all of the subhalos are suitable hosts of a galaxy with the present day luminosity of Fornax. We have made this assumption in order to analyze a large statistical sample of well-resolved Fornax host candidates in N-body simulations that do not account for the effects of baryons on the subhalo evolution. High resolution Local Group simulations with baryons have been recently undertaken~\citep{Sawala_etal14,Sawala:2014xka}, though at present the statistical sample of subhalos is smaller in these simulations than the corresponding sample from N-body simulations. For this reason, we use semi-analytic models for the luminosities of our Fornax subhalo host candidates, and leave to a future study the analysis of the simulations that include the baryons. 

Our subhalo luminosities are motivated by the results of several studies. First, we consider the results of~\cite{Sawala_etal14}, who utilize hydrodynamic simulations to predict the luminosity of subhalos before they fall into the Milky Way halo and at $z=0$. For subhalos with the present day luminosity of Fornax, Sawala et al. find that subhalos with the luminosity of Fornax have a total mass in the range $\sim 2-10 \times 10^9$ M$_\odot$, and a total peak halo mass in the range $\sim 7-20 \times 10^9$ M$_\odot$. The high end of this mass range is consistent with previous semi-analytic models~\citep{Cooper:2009kx} and with the predictions from the abundance matching method~\citep{Behroozi_etal13}. For dSphs fainter than Fornax, the predicted range of subhalo masses for a fixed luminosity is larger, and also there are stronger deviations from the predictions of abundance matching and that of~\cite{Sawala_etal14}. A well-known caveat, however, is that the abundance matching technique is not calibrated at both sub-galactic scales and in non-CDM cosmologies. Therefore results in these regimes depend on extrapolating the existing models and assuming the average galaxy formation history holds for non-CDM based models. 

In order to determine the relevant peak subhalo mass range for Fornax, we first take the lower bound on the stellar mass of Fornax to be $M_{*}\gsim$1$\times10^{7}M_{\odot}$. The lower bound is derived from the 1 $\sigma$ lower bound on the Fornax V-band luminosity of $L_V =$ 1.7 $\pm ^{0.5}_{0.4}\times 10^7 L_{\odot}$, and assuming a stellar mass-to-light ratio $M_{*}$/L = 0.8 ($M_{*}\sim1.3\times 10^7\times 0.8 M_{\odot}$ for the lower bound). For this adopted stellar mass, we then select Fornax candidates by their peak mass that predict the same stellar mass range using the abundance matching description in~\citet{Behroozi_etal13}. In Figure~\ref{fig:mpeak} we show the distribution of peak subhalo masses for these Fornax candidates. These results indicate that while subhalos that are consistent with the Fornax kinematics have subhalo peak masses in the range $M_{sub} \sim 10^8$-$10^{10} M_{\odot}$, only a handful at the very massive end predict the right luminosity. For example from the right panel in Figure~\ref{fig:mpeak} we find that 2 out of 124 candidates from the six Aquarius simulations are consistent with the Fornax luminosity. Interestingly, these two candidates come from the Aquarius A and B simulations, which are, respectively, the most massive and the least massive halos among all six Aquarius simulations. This would indicate that the likelihood of generating a Fornax candidate may be only weakly correlated with the galactic halo mass.

\begin{figure*}
\includegraphics[height=6.5cm]{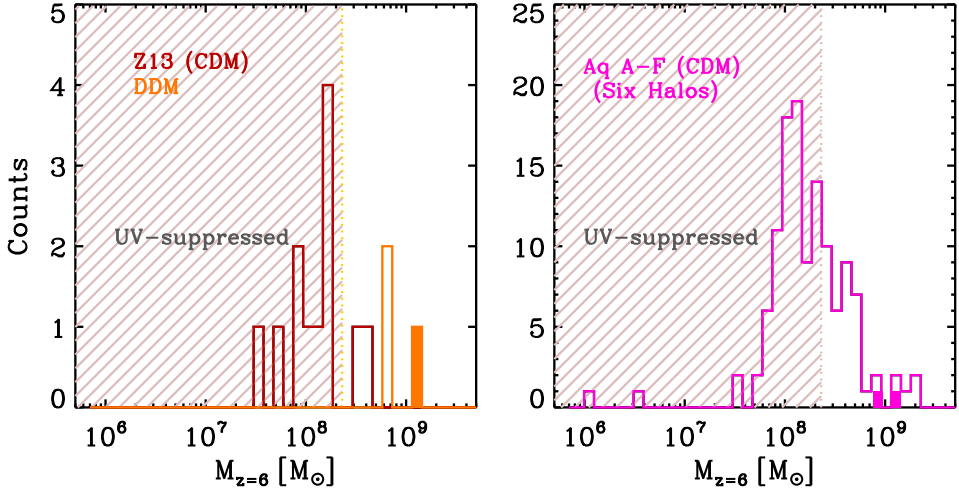}
\caption{ 
The distribution of the progenitor mass at z=6 for the Fornax candidates. The color scheme in the left panel is the same as Figure~\ref{fig:param} showing DDM and Z13 simulation candidates, and right panel shows all the candidates from all six Aquarius halos.  We note that there is only one galactic halo realization in the left panel, while there are six in the left panel. The filled histograms indicate the subhalos with luminosity that match Fornax at z=0. The shaded areas indicate the progenitor mass range with baryon content subject to UV background suppression~\citep{Okamoto_etal08}. 
}
\label{fig:mz6}
\end{figure*}

For comparison, in Figure~\ref{fig:mpeak} we show the Fornax candidates from the WDM and DDM simulations, as well as their CDM counterparts. Though the sample of host subhalos is smaller than in the case of CDM, there is a higher probability for WDM and DDM Fornax candidates to reside in more massive subhalos. Therefore, the WDM and DDM candidates are more likely to match the bright luminosity of Fornax than the CDM candidates. In each case, we find 1-2 candidates that match the kinematics and luminosity, out of 3-7 candidates that just match the kinematics.  Note again the caveat that the relationship between the stellar mass and subhalo progenitor mass may differ from this relation which is derived from CDM simulations. 

With candidate host subhalos identified we are now in position to study their evolution and the masses of their progenitors near the redshift of reionization. The effect of reionization on suppressing star formation in small halos shows that halos below a few $10^{8} M_{\odot}$ likely have no stars due to the UV-background suppression. This implies that any subhalo with a progenitor less massive would be unlikely to host a visible galaxy today~\citep{Okamoto_etal08, Sawala_etal14}. 

In Figure~\ref{fig:mz6} we show the subhalo progenitor mass at z=6 for the Fornax candidates. According to the criteria of~\citet{Okamoto_etal08}, more than half of the Fornax candidate in Aquarius would not host any stars. The subhalos that match both the Fornax kinematic and photometry data as well as the luminosity are on the massive end of the distribution.

In the left panel of Figure~\ref{fig:mz6} we show the case for the DDM simulation and its CDM counterpart. We do not include WDM simulations here because their structure formation and galaxy formation at high redshift are expected to be significantly different than in CDM. Also for the WDM models that we considered here, majority of the subhalo progenitors are not yet formed or identified by halo finders at z=6. However, for the DDM model with a decay lifetime of $10$ Gyr, it is safe to assume a similar reionization history as CDM. Since at $z=6$ less than about 10$\%$ of the dark matter has decayed with a small recoil kick velocity of $20$ km/s, the structure growth of both the dark matter and baryon components should be similar to CDM. Again the majority of the CDM subhalos should be UV-suppressed, and all three DDM subhalos are above the threshold, with one object that matches the Fornax luminosity. 

\subsection{Mass Assembly Histories}
\label{subsection: assembly}
We now move on to discuss the mass assembly histories of our Fornax candidates. The left panel of Figure~\ref{fig:mz} shows the subhalo mass as a function of redshift for all 124 of our Fornax candidates from all six Aquarius halos. The middle panel shows the corresponding candidates from the WDM model with 2.3 keV mass and the right panel shows the candidates from DDM simulations. From the discussion above, we identify the subhalos that match the Fornax luminosity, and also determine an approximate infall time for those subhalos that match both the Fornax luminosity and kinematics. From Aquarius simulations, we can see that the candidates that match the kinematics and luminosity, which are shown in blue lines rather than the gray lines for candidates only fit stellar kinematics, also have the largest ratio of M($z_{peak})/M(z=0)$, with typical peak values for this ratio of $\sim$ 30$-$60 at $z\sim$2. For WDM, subhalos that match the kinematics and luminosity also show a large peak value at $z \sim 1.5$. On the other hand the corresponding DDM subhalo shows a relatively low peak value of M($z_{peak})/M(z=0) \sim$ 3 at $z~\lsim 1$. 

From the results presented in Figure~\ref{fig:mz}, we can identify three important trends and predictions regarding tidal stripping, infall times, and star formation histories.

\subsubsection{Tidal Stripping}
\label{Tidal stripping}
Using the above results in CDM we have a clear prediction that Fornax has lost a significant amount of dark matter mass after it falls into the Milky Way.  Does this translate into a more specific observational signature? One possibility is that this tidal stripping of dark matter is manifest in the tidal stripping of the stars. However, since our models only directly include dark matter, it is not clear whether we should expect stellar tidal tails to be observable. It is possible that the stellar components are embedded deeply in the center of the subhalo potential, so that the dark matter is mostly stripped and the stars left unstripped. For example, ~\cite{Watson_etal12} shows that if subhalos experience substantial dark matter mass loss before mass is lost within the galaxy, this explains how satellite galaxies lose stellar mass and contribute to ``intrahalo light" (IHL).  

\begin{figure*}
\includegraphics[height=7.0cm]{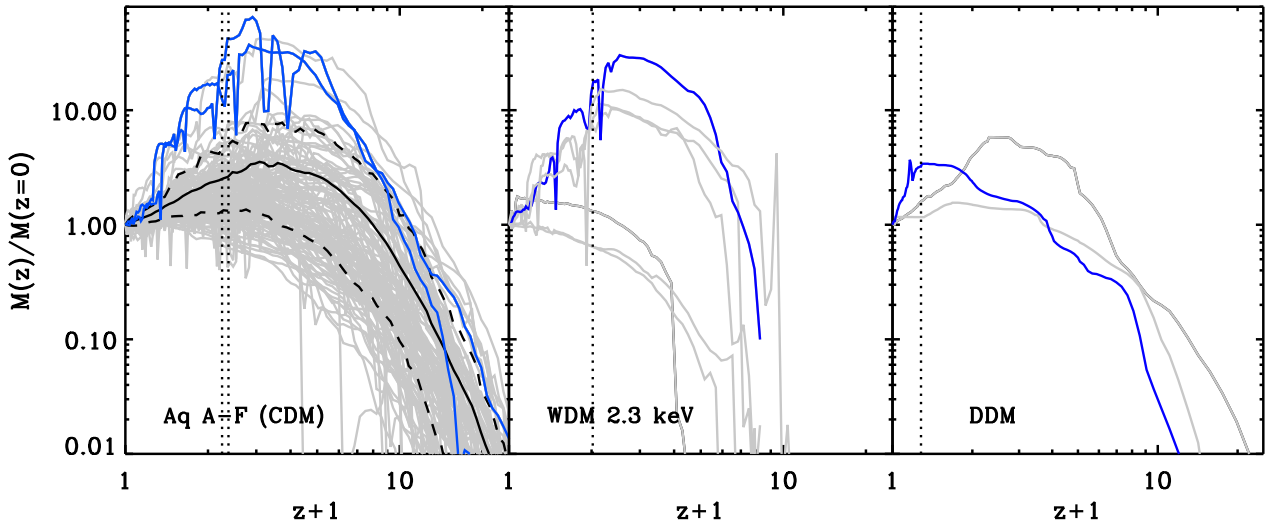}
\caption{ 
Mass assembly histories for the Fornax good-fit subhalos. The gray lines show the fits with luminosity dimmer than Fornax using abundance matching technique from \citet{Behroozi_etal13}, and the blue lines show those with luminosity that matches the Fornax luminosity. The vertical dotted lines show the infall time of the blue curves. In the left panel the dashed black lines include 68$\%$ of the Aquarius simulation fits, and the black solid lines show the mean value.
}
\label{fig:mz}
\end{figure*}

For comparison the Fornax candidates in WDM, shown in the middle panel of Figure~\ref{fig:mz}, are similar to CDM in that both require large $M_{peak}/M(z=0)$. In DDM, as shown in the right panel of Figure~\ref{fig:mz}, the ratio is much smaller, $M_{peak}/M(z=0)~\sim 3$. These results hint at the interesting possibility that hosts of Fornax have undergone different degrees of tidal stripping in different dark matter models. Thus with future larger statistical samples of non-CDM models, as well as the inclusion of baryons in the simulation, the degree of tidal stripping of Milky Way-satellites may provide a new test of dark matter models.

\subsubsection{Infall Time}
\label{Infall}
From the peaks in the evolution of the subhalo masses in Figure~\ref{fig:mz}, we can identify the infall times of the subhalos into the Milky Way halo. In CDM, the subhalos that match the Fornax kinematics and luminosity have infall redshifts of $z=1.25$ and $z=1.37$, corresponding to lookback times of $\sim 9$ Gyr. For the 2.3 keV WDM model the infall redshift is $z=1.02$, corresponding to a lookback time of $\sim$ 8 Gyr. For DDM model the infall redshift is at $z=0.28$, corresponding to a lookback time of $\sim$ 3 Gyr. So the DDM host subhalo is predicted to fall into the Milky Way more recently than the CDM and WDM candidates. 

It is informative to compare the lookback times that we deduce to previous estimates of the infall times of Milky Way satellites.~\cite{Rocha_etal12b} estimate infall times using subhalo Galactocentric positions and orbital motions from Via Lactea II simulation. Using these criteria Fornax is most likely to have fallen into the Milky Way halo $\sim 5-9$ Gyr ago. So even though our criteria for identifying Fornax subhalo candidates in CDM simulations are different from those in~\cite{Rocha_etal12b}, the infall times are in good agreement. We note that the Aquarius simulations that we utilize and our stricter matching criteria using luminosity and stellar kinematic data allows for a larger range of merger histories than considered in~\cite{Rocha_etal12b}, who utilize a single Milky Way halo. 
  
\subsubsection{Star-Formation Quenching Mechanism }
\label{Quenching}
The quenching of star formation in Milky Way dSphs may be related to their infall times. For example ram-pressure stripping can remove the cold gas at the center of the satellites as a result of the high-speed interaction with the hot gas halo of the host halo \citep{Gunn_etal1972}. Ram pressure stripping has been invoked as the quenching mechanism for Milky Way and M31 dwarf satellites with $M_\star~\lsim~10^8 M_{\odot}$ with an extremely short quenching timescale $\sim$ 2 Gyr~\citep{Fillingham_etal15}. 

\begin{figure}
\includegraphics[height=5.8cm]{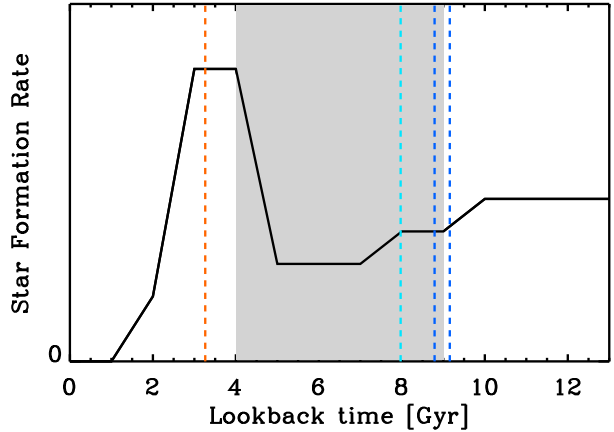}
\caption{ 
Qualitative Fornax star-formation histories (black solid lines, from \citet{Coleman_etal08}) compared to estimated infall times from \citet{Rocha_etal12b} using Galactocentric positions and orbital motions (grey filled area) and our results (vertical dashed lines). The two navy dashed lines indicate the infall time for the two CDM candidates from the Aquarius simulations that match both the Fornax luminosity and stellar kinematics. The aqua line is for the WDM 2.3 keV candidate, and the orange line is for the DDM candidate. Notice the DDM candidate infall time is significantly lower than the \citet{Rocha_etal12b} prediction and our CDM and WDM candidates, and also it happens at the time when the Fornax star formation rate starts to drop after a burst 3-4 Gyr ago. All others predict infall times before the enhanced star formation event.
}
\label{fig:sf}
\end{figure}

As shown in Figure~\ref{fig:sf}, Fornax is observed to have an enhanced star formation activity $\sim$ 3-4 Gyr ago and quenching $\sim$ 2 Gyr ago \citep{Coleman_etal08}. If we match these timescales with the infall times that we determined above, the enhanced star formation event is most consistent with the infall time of our DDM subhalo candidate. The grey shaded area shows the possible Fornax infall time range from ~\citet{Rocha_etal12b}. This covers a large infall time range including those predicted by the CDM and WDM simulations. However, it is interesting to note that other recent Fornax star formation history studies either indicate similar peak at $\sim$ 4 Gyr~\citep{deBoer_etal12} or draw different conclusions with a star formation peak at $\sim$ 8 Gyr~\citep{del_Pino_etal13}. Deep photometric studies of Fornax will help resolve this discrepancy and provide more insight on the connection of the Fornax star formation history with its infall time.

To explain the Fornax star formation history in the context of the earlier subhalo infall times predicted in our CDM and WDM models, we can consider a scenario in which the Fornax star formation was quenched when it merged into Milky Way halo $\sim$ 9 Gyr ago. A close encounter with the Milky Way at the last perigalactic passage would then trigger star formation. This mechanism has been invoked to explain the star formation history of Leo I~\citep{Mateo_etal08}, which had a burst in its star formation $\sim 3-4$ Gyr ago. Indeed proper motion and line-of-sight velocity measurements indicate that Leo I is likely on a fairly eccentric, nearly unbound orbit~\citep{Sohn_etal13}. In contrast with Leo I, Fornax likely has a much less eccentric orbit~\citep{Lux_etal10,Piatek_etal07}, so a much closer encounter with the Milky Way may be more difficult to invoke in the case of Fornax. More precise measurement of orbital velocity and detailed orbit reconstruction will shed more light on this issue. 

\section{Conclusions and Discussion}
\label{section:conclusion}

We have fit the kinematics and photometry of the Fornax dSph to subhalo potentials predicted by numerical simulations of different dark matter models. 
CDM subhalo candidates are typically in the mass range of $10^8-10^9 M_{\odot}$. WDM and DDM models predict on average larger mass subhalo hosts of Fornax. 

The diversities of subhalo candidate properties in different dark matter models and the different time evolution of dark matter free-streaming effects lead to diversities in subhalo formation history, and also likely their stellar formation history. With this in mind we utilize simulation merger trees to investigate the formation history of the subhalos that match the Fornax kinematics. We implement simple models to match the Fornax luminosity to dark matter halo mass in order to understand possible star formation histories of these Fornax subhalo candidates. Under the assumption that the mass of the most massive progenitor is correlated with the current stellar content in each galaxy, we derive current stellar masses for each subhalo.

For CDM subhalos, our best candidates require the ratio between the peak mass and current mass to be $\gsim$ 10. This suggests that these systems experience significant tidal stripping after they fall into Galactic halo and are subject to substantial dark matter mass loss. It also implies that their infall times are $\gsim$ 9 Gyr ago, so that they fall well inside the galactic halo where the tidal stripping is efficient. Interestingly, similar findings also have been pointed out by~\cite{Cooper:2009kx} using Aquarius simulations with a ``particle tagging" technique to simulate Galactic stellar halos. Using Fornax and Carina as two examples, these authors find that they can only match their observed dwarf galaxy surface brightness and velocity dispersion profiles simultaneously by choosing model satellites that have suffered substantial tidal stripping. 

Our WDM subhalo candidates show only mild deviation from their CDM counterparts, in that they have similar peak mass to current mass ratio and infall time. A caveat to this analysis is the real stellar content might be different, and full hydrodynamic simulations are needed to calibrate galaxy formation process in WDM. For the DDM candidates, using similar assumptions our subhalo candidates have relatively small peak mass/current mass ratio ($\sim$ 3) and they are more massive at $z=0$ than the CDM candidates. Therefore the DDM candidate only experiences mild mass loss since merging into the Galactic halo approximately 3 Gyr ago. 

We note that our results are not yet able predict the existence of visible stellar tidal tails. Since galaxies are embedded deep in the center of the halo potential, most of the tidal stripping effects are expected to be on the DM component. Determining whether or not stars are stripped out requires N-body simulations with star particles.~\cite{Cooper:2009kx} have shown that 20$\%$ of the stellar mass and 2$\%$ of DM mass of their Fornax candidate remains at present time. Also recent work from \cite{Battaglia_etal15} have examined the tidal effects on Fornax along its possible orbits. Current and forthcoming wide field deep imaging surveys may be sensitive to faint stellar tidal tails and can provide an observational test of our models. ~\citet{Bate_etal15} use data from VLT Survey Telescope (VST) ATLAS Survey to study a region of 25 square degrees centered on Fornax. They have excluded a shell structure outside of Fornax's tidal radius that was reported from a previous study \citep{Coleman_etal05}. In the near future the Dark Energy Survey (DES) will cover a wide area around Fornax with much improved imaging depth, and possibly provide a test of the tidal stripping hypothesis. 

We find that most of the subhalo candidates are subject to the cosmic reionization UV-background so that their star formation is highly suppressed. In our sample more than half of the CDM candidates are likely not visible at all due to reionization. For the WDM model the reionization history is more uncertain because high-redshift structure formation is significantly delayed. However, for DDM the early structure formation follows CDM until late times when the decay process becomes significant, so a similar reionization history can be applied. Most of our DDM candidates are not affected by the UV-background since they are already massive at z=6. This is largely because for the same Fornax kinematic data, DDM candidates tend to be more massive because they have less concentrated DM profiles than their CDM counterparts.

The difference in Fornax infall times in the different models may imply a different star formation quenching mechanism if environmental effects play an important role in dSph formation processes. From the observed Fornax star formation history, there is an enhanced star formation event at 3-4 Gyr ago and it is quenched $\sim$ 2 Gyr ago~\citep{Coleman_etal08}. For those subhalo candidates that have infall times well before the star formation peak, star formation may be triggered by a close passage to the Milky Way. For those that have infall times after the star formation peak, the infall into Galactic halo may cause rapid gas loss due to ram-pressure stripping. More precise predictions can be made by detailed Fornax orbital motion reconstruction from its proper motion and line-of-sight velocity measurement, which may be possible with the GAIA satellite. Larger sets of simulations that sample more possible halo formation histories in different dark matter scenarios will also help to confirm our results.

%
%
\section*{Acknowledgments}
We would like to thank Mike Boylan-Kolchin, Andrew Cooper, Yao-Yuan Mao, Julio Navarro, and Till Sawala for useful discussions. M.-Y. W. and L.E.S. acknowledge support from NSF grant PHY-1522717. This work is part of the D-ITP consortium, a programme of the Netherlands Organization for Scientific Research (NWO) that is funded by the Dutch Ministry of Education, Culture and Science (OCW). This work was supported by the Science and Technology Facilities Council (grant number ST/F001166/1) and the European Research Council (grant numbers GA 267291 "Cosmiway"). It used the DiRAC Data Centric system at Durham University, operated by the Institute for Computational Cosmology on behalf of the STFC DiRAC HPC Facility (www.dirac.ac.uk). This equipment was funded by BIS National E-infrastructure capital grant ST/K00042X/1, STFC capital grant ST/H008519/1, and STFC DiRAC Operations grant ST/K003267/1 and Durham University. DiRAC is part of the National E-Infrastructure. The work of ARZ was funded in part by the Pittsburgh Particle Physics, Astrophysics, and Cosmology Center (Pitt PACC) at the University of Pittsburgh.

\bibliography{fornax_dyn}

\appendix
\section{Resolution Tests and Corrections}
\label{Resolution}

\begin{figure*}
\includegraphics[height=7.6cm]{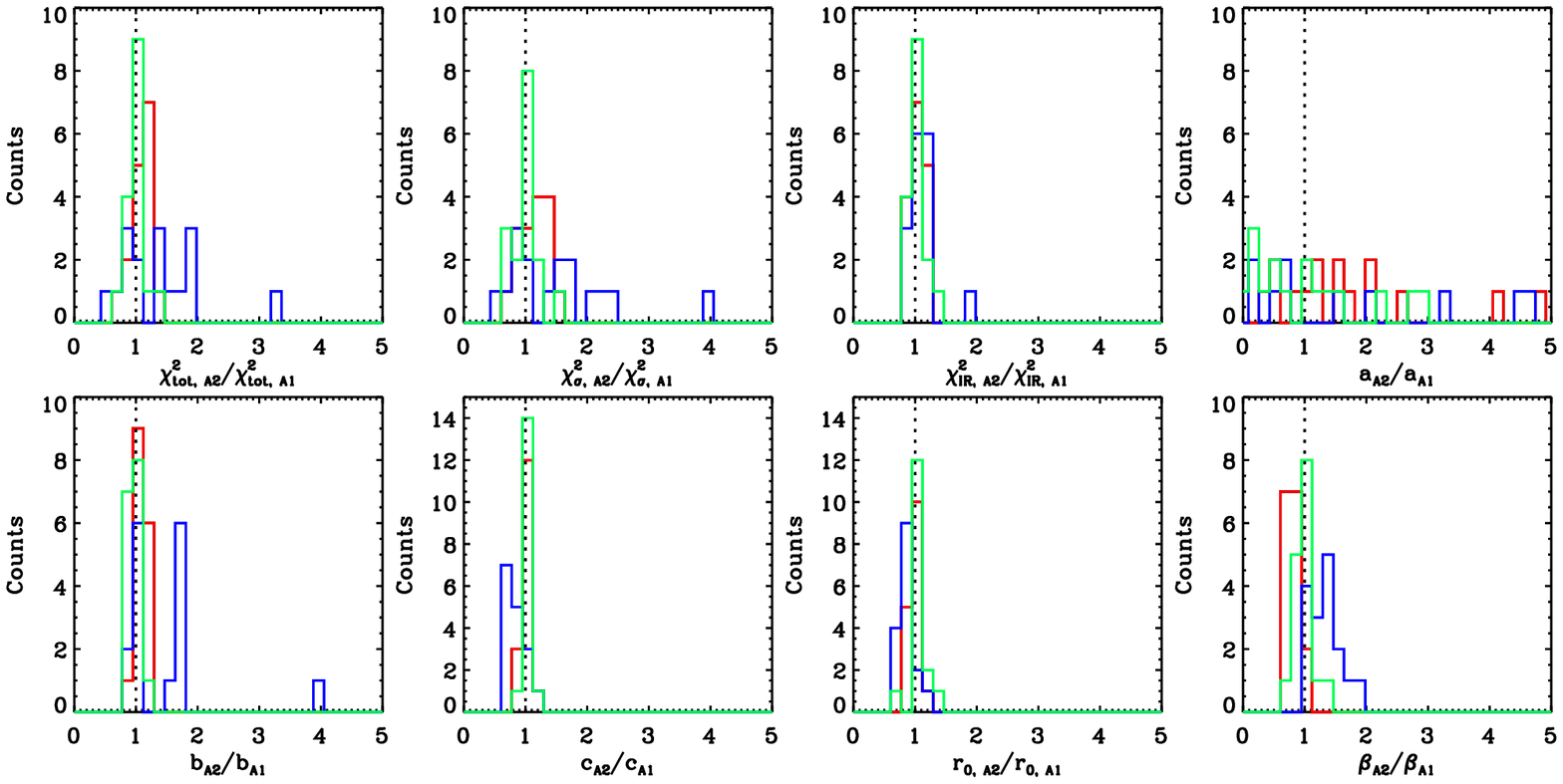}
\caption{ 
The distribution of the fitting results for Aquarius A-2 simulations with different softening scale correction methods : no correction (red histogram), blue histogram with $r_{inner}$= 300 pc, and  green histogram $r_{inner}$=200 pc. Here $r_{inner}$ is within what radius a Einasto profile fit is used for accumulated mass profile. The x-axes show the ratio of fits to their Aquarius level 1 counterparts.
}
\label{fig:cor_param}
\end{figure*}

As pointed out in \cite{Boylan-Kolchin_etal12} where they used Aquarius simulations to compare dynamic properties of simulated galactic subhalos with observed Milky Way satellites, the effect of force softening will introduce errors to the subhalo potentials in the central region. The reason is that the force softening scales of our simulations range from $\epsilon$ = 65.8 - 72.0 pc. The effect of force softening will reduce the density on scales of $\sim$ 3 $\epsilon$. However we need to work with a cumulative mass profile that will carry the effect to larger scale. In order to estimate and further correct for this effect, here we utilize the Aquarius A level 1 simulation, which has the same halo realization as the Aquarius A level 2 simulation but with much smaller force softening length ($\epsilon$ = 20 pc) and particle mass ($m_p$ = 1.712 $\times 10^3$ $M_{\odot}$), to test different correction methods to reconstruct the mass distribution at subhalo center.

The correction method adopted here is similar to the one in \cite{Boylan-Kolchin_etal12}. We have tried out a few different fitting ranges and correction ranges to find the one that gives the best results. The procedure is as follows. We fit the subhalo density profiles with an Einasto profile, which has the following form:
\beq
\rho(r)=\rho_{-2} \, \textrm{exp}\Big(-{2 \over \alpha} \Big[ \Big( {r \over r_{-2}}\Big) ^{\alpha} -1\Big] \Big). 
\eeq
The fitting range is [$r_{inner}$ ,$r_{upper}$], where $r_{upper}$ is the smaller of 3 kpc or 1.5 $r_{max}$ and we try a few different values for $r_{inner}$. We then replace the mass within $r_{inner}$ with the accumulated mass generated from the Einasto profile fit:

\beq
M(<r)={4\pi r_{-2}^3 \, \rho_{-2} \over \alpha} \textrm{exp}\Big({3 \, \textrm{ln} \alpha+2-\textrm{ln}8 \over \alpha} \Big) \gamma \Big[ {3\over \alpha},{2\over \alpha}\Big( {r \over r_{-2}}\Big)^{\alpha} \Big]
\eeq
and use the mass profile from simulations outside of the $r_{inner}$.

In Figure~\ref{fig:cor_param} we compare the Fornax data fitting results using different corrections with different $r_{inner}$ on Aquarius level 2 subhalos to their Aquarius level 1 counterparts. Here we show three different $r_{inner}$ choices: no correction ($r_{inner}$ = $\infty$, red histograms in Figure~\ref{fig:cor_param}), $r_{inner}$= 300 pc (close to choice of 291 pc in \cite{Boylan-Kolchin_etal12}, blue histograms in Figure~\ref{fig:cor_param}), and 200 pc (green histograms in Figure~\ref{fig:cor_param}). We take the ratio of the fitting parameter value from Aquarius level 2 subhalos with their level 1 counterparts. So the ideal case will be they lie very close to 1, indicating that the corrections faithfully recover the mass distribution in higher resolution simulation runs.

We can see that some of the parameters are not affected much by the correction. Examples are $c$ and $r_0$, since they are related to stellar distribution in the outer region (for example, $r_0 \sim$ 1 kpc) and are not sensitive to the correction in the inner region ($\lsim$ 300 pc). The $\chi^2$ of the photometry data fit also is not affected significantly, since most of the photometry data points are distributed outside $\lsim$ 300 pc. The worst of all parameters is $a$, and the amount of deviation seems to be large no matter which correction we apply. The reason is that the data do not provide good constraints on the parameter $a$. The uncertainty of the MCMC result for $a$ is usually very large and highly correlated with other parameters such as the parameter that determines the normalization of the photometry data. So the wide distribution of the parameter $a$ for different corrections is likely not due to the way the dark matter potential is corrected, but due to the large uncertainties from the poor constraints and parameter correlations. From the overall behavior of these three different correction methods, we can see that the green histograms, for which the Einasto profile correction is applied within 200 pc, show the best agreement with Aquarius level 1 counterparts. 

\begin{figure}
\includegraphics[height=3.8cm]{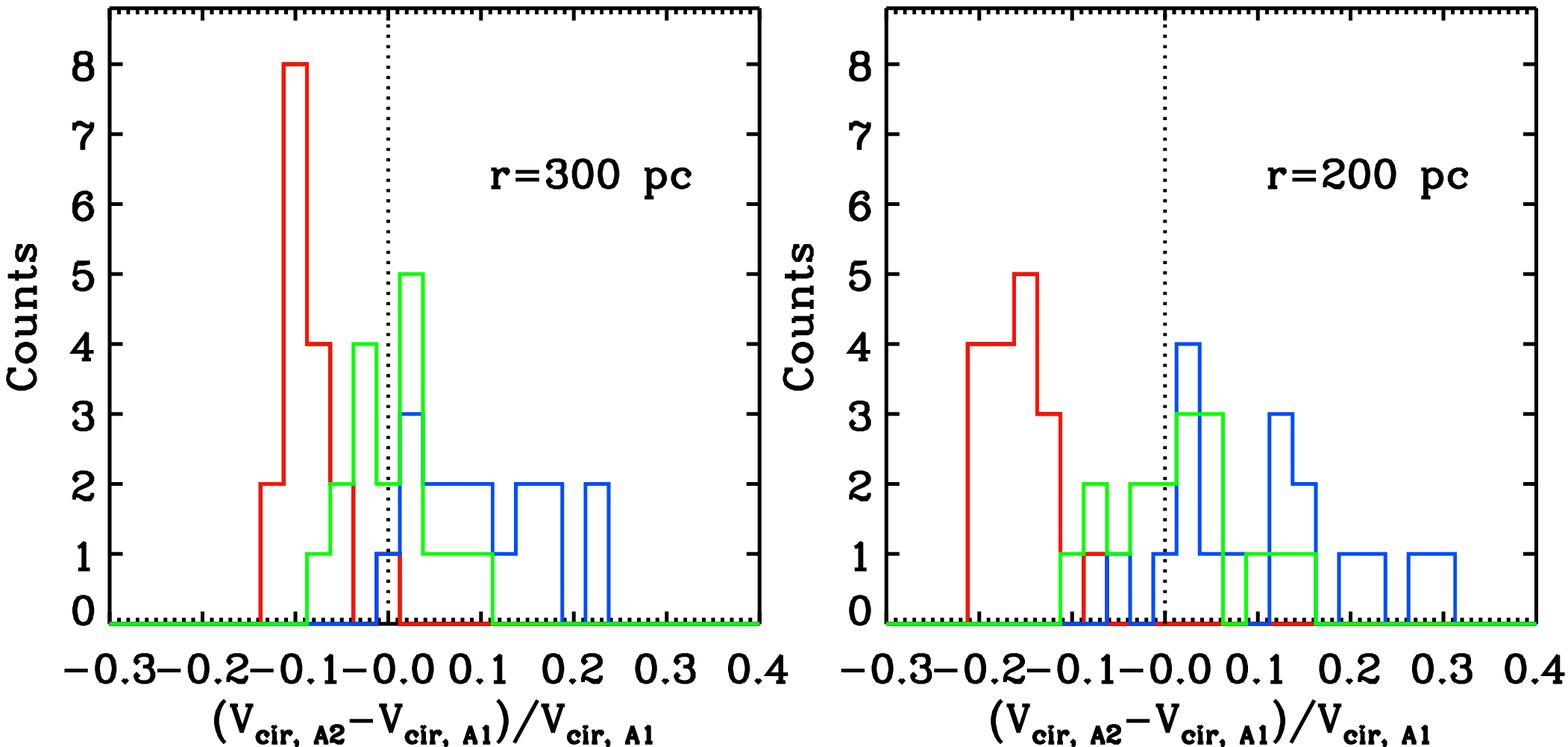}
\caption{ 
The distribution of circular velocity ($V_{cir}$) fractional difference for different Aquarius A resolution simulations and different softening scale correction methods. The x-axis shows the fractional difference of the $V_{cir}$. The color scheme is the same as Figure~\ref{fig:cor_param}. 
}
\label{fig:vcir_diff}
\end{figure}

In Figure~\ref{fig:vcir_diff} we show the fractional difference in circular velocity between Aquarius level 1 subhalos and their level 2 counterparts after corrections at two radii: r=200 and 300 pc. At 300 pc the difference is within 20$\%$ and at 200 pc the difference is within 30$\%$ for all correction methods. However, before applying any correction, the difference is negative (red histograms) and for the correction applied within 300 pc (blue histograms) the difference is biased toward positive values. The green histogram, which is for the correction within 200 pc, is peaked at zero and with difference within 10-15$\%$. This indicates that before any corrections the central dark matter potentials ($V_{cir} = \sqrt{GM(r<)/r}$) are reduced due to softening scale effects for Aquarius level 2 simulations, and the suppression is up to about 20$\%$. However, if we apply the correction method suggested in \cite{Boylan-Kolchin_etal12}, we can "over-correct" the potentials up to 20-30$\%$ at 200-300 pc. Therefore our tests show that a mild correction within 200 pc with the fitted Einasto profile mass will generate the best results for removing the softening scale effects. We thus adopt the correction all the simulations in this study.

\section{Effects of WMAP 1 $\&$ WMAP 7 cosmological parameters}
\label{w1w7}

\begin{figure*}
\includegraphics[height=5.0cm]{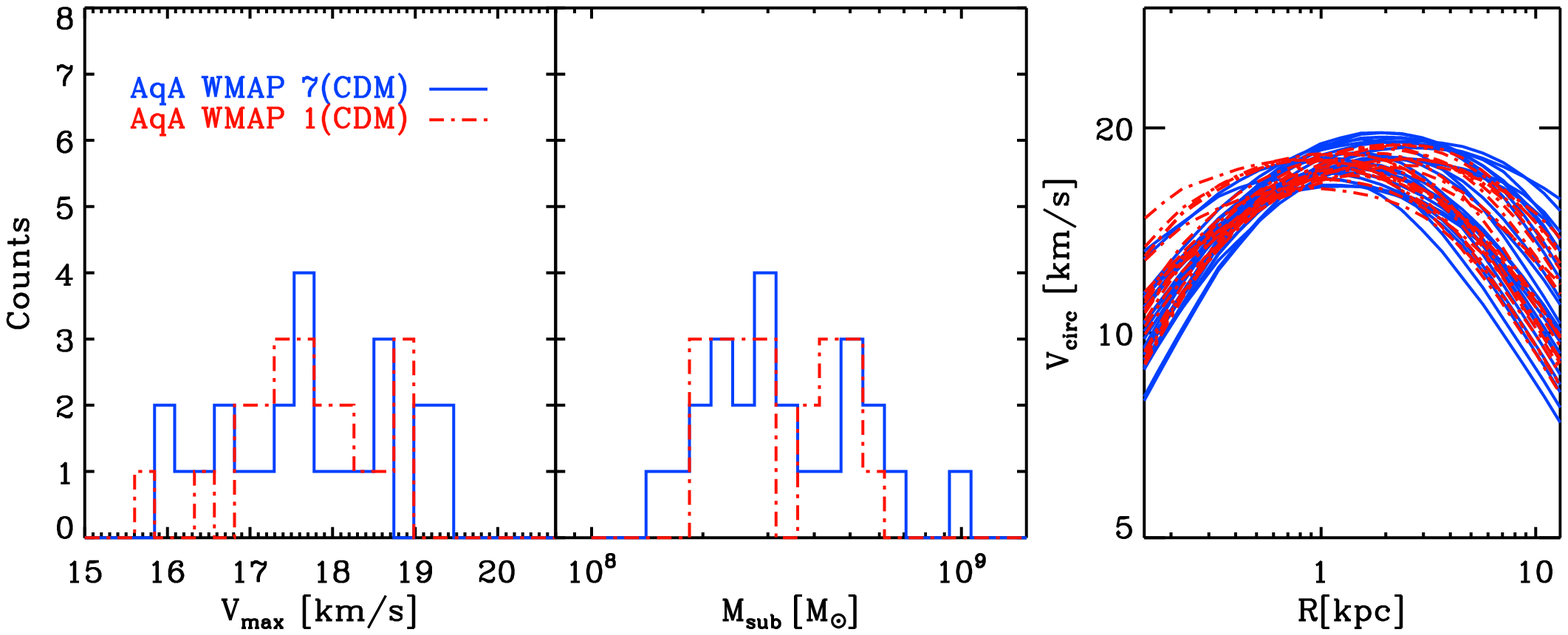}
\caption{ 
Subhalo properties at z=0 : maximum circular velocity ($V_{max}$, left panel), and subhalo mass ($M_{sub}$, middle panel), and circular velocity profiles (right panel) for the Fornax good-fits from AqA WMAP 1 (red dash-dotted lines) $\&$ AqA WMAP 7 (blue solid lines) simulations. Here we only show those we can find counterparts in both simulations.
}
\label{fig:halo_w1w7}
\end{figure*}

Here we test the effects of different cosmological parameters on our study. Previous studies (e.g. \citet{zentner_bullock03, Polisensky_etal14}) have shown that different cosmological parameter values, especially $\sigma_8$ and $n_s$, have non-negligible effect to the substructure density profiles. The simulation sets used in our study have adopted different cosmological parameter values. We compare  the Fornax Jeans equation fitting results from the Aquarius A Level-2 simulation that is re-simulated with a WMAP 7 cosmology (Aq-A2-w7) with those from the original Aquarius A Level-2 simulation that is simulated with a WMAP 7 cosmology (Aq-A2). One of the most significant differences of their choice of parameters lies in the difference in $\sigma_8$ ($\sigma_8$ = 0.9 for Aq-A2 while $\sigma_8$ = 0.81 for Aq-A2-w7). In the following we will compare the fitting results and discuss the impacts due to the differences in cosmological parameters. 

In Figure~\ref{fig:halo_w1w7} we show the present-time subhalo properties of the Fornax candidates and their circular velocity curves. Those are candidates that fit Fornax kinematics but not necessarily fit the Fornax luminosity. In both simulations we find a comparable number of good-fits. Here we only show those we can find counterparts in both simulations. Their maximum circular velocities ($V_{max}$) and subhalo mass ($M_{sub}$) span very similar range. However, it is shown in the right panel in Figure~\ref{fig:halo_w1w7} that those subhalos from Aq-A2 are slightly more concentrated at the center than those from Aq-A2-w7. As discussed in~\citet{Polisensky_etal14}, halos in simulations with higher $\sigma_8$ form earlier and thus are more concentrated. However, we can see from the distribution of candidate subhalo properties that the changes due to the differences in WMAP 1 and WMAP 7 cosmology are much smaller than the difference due to different dark matter properties. When we compare Figure~\ref{fig:halo_w1w7} with Figure~\ref{fig:param}, we can see that WDM and DDM simulations generate subhalos that exhibit very different properties that occupy a completely different subhalo property region. It is thus clear that high-quality Fornax data have provided good constraints on subhalo present-time properties.

\begin{figure}
\includegraphics[height=4.2cm]{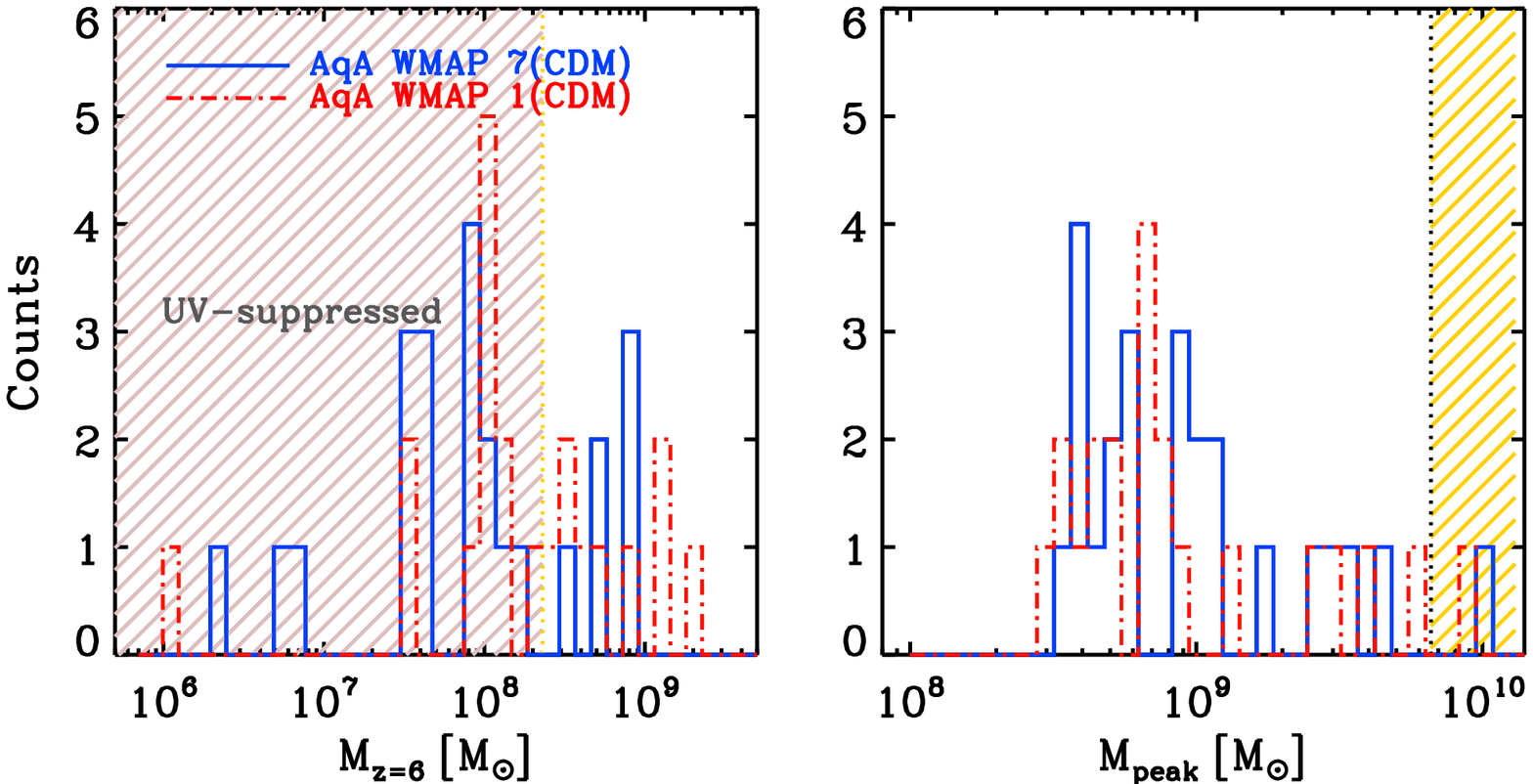}
\caption{ 
The  $M_{peak}$ and mass at z=6 ($M_{z=6}$) distribution of the Fornax good-fits from the AqA WMAP 1 (blue solid histogram) and AqA WMAP 7 (red dash-dotted histogram) simulations. The shaded area in the left panel indicates the progenitor mass range with baryon content that is subject to the UV background suppression. The yellow shaded area in the right panel indicates the region where the luminosity of these objects are consistent with the Fornax luminosity using abundance matching methods from \citet{Behroozi_etal13}.
}
\label{fig:mp_w1w7}
\end{figure}

We also investigate the differences in halo formation history. In Figure~\ref{fig:mp_w1w7} we check the distribution of $M_{peak}$ and mass at z=6 of Fornax candidates from these two simulations. We find that the difference is mild and does not affect our conclusion. For example, for both simulations more than half of the candidates may have suppressed star-formation due to the UV-background from reionization. Also for both simulations only one candidate matches the Fornax luminosity while others are much dimmer. Thus we conclude that the difference in WMAP 1 and WMAP 7 is mild and it has limited effects on both present-time properties and halo formation history. The differences due to dark matter properties, which is the focus of our study, exhibit much stronger effects.

\end{document}